\documentclass[twocolumn,english,prb,showpacs]{revtex4}
\usepackage[T1]{fontenc}
\usepackage{amssymb}
\usepackage{graphicx}
\usepackage{amsmath,color}
\usepackage{natbib,hyperref}
\usepackage[squaren,Gray]{SIunits}
\usepackage{multirow,bigdelim}

\usepackage{lipsum}

\usepackage{color}

\begin{document}


\title{Structure, Energy, and Thermal Transport Properties of Si-SiO$_2$ Nanostructures using an {\it Ab initio} based Parameterization of a Charge-Optimized Many-Body Forcefield }

\author{Arthur France-Lanord$^{a),b)}$}
\email[]{afl@materialsdesign.com}

\author{Patrick Soukiassian$^{b)}$}
\author{Christian Glattli$^{b)}$}
\author{Erich Wimmer$^{a)}$}

\affiliation{$a)$ Materials Design SARL, 92120 Montrouge, France}
\affiliation{$b)$ Commissariat {\`a} l'Energie Atomique et aux Energies Alternatives, DSM-IRAMIS-SPEC and UMR CNRS 3680, Saclay, 91191 Gif-sur-Yvette Cedex, France}


\date{\today}

\begin{abstract}
In an effort to extend the reach of current {\it ab initio} calculations to simulations requiring millions of configurations for complex systems such as heterostructures, we have parameterized the third-generation Charge Optimized Many-Body (COMB3) potential using solely {\it ab initio} total energies, forces, and stress tensors as input. The quality and the predictive power of the new forcefield is assessed by computing properties including the cohesive energy and density of SiO$_2$ polymorphs, surface energies of alpha-quartz, and phonon densities of states of crystalline and amorphous phases of SiO$_2$. Comparison with data from experiments, {\it ab initio} calculations, and molecular dynamics simulations using published forcefields including BKS (van Beest, Kramer, and van Santen), ReaxFF, and COMB2 demonstrate an overall improvement of the new parameterization. The computed temperature dependence of the thermal conductivity of crystalline alpha-quartz and the Kapitza resistance of the interface between crystalline Si$(001)$ and amorphous silica are in excellent agreement with experiment, setting the stage for simulations of complex nanoscale heterostructures.\end{abstract}


\maketitle

\section{Introduction}

Understanding and quantitative prediction of structural, vibrational, and thermal properties of nanoscale heterosystems requires a detailed energetic description based on atomic structure. Despite the staggering advances in computer hardware and the progress in computational methods, the limitations of current {\it ab initio} calculations make it necessary to use interatomic potentials or forcefields to simulate millions of configurations of systems with many thousands of atoms. Forcefield methods are well developed for homogenous systems such as pure metals, ionic compounds, liquids, and polymers. While a wealth of forcefields for pure SiO$_2$ and Si have been reported\cite{Hill1998, Sanders1984, Tsuneyuki1988, vanBeest1990, Vashishta1990, Tangney2002, Tersoff1986, Stillinger1985}, the development and parameterization of forcefields for heterosystems such as semiconductor/oxide interfaces remains a formidable challenge due to effects including charge transfer and changes in the bonding topology. 

In this context, silicon and its oxides play a prototypical role, especially as they are omnipresent in nanoscale electronic devices. While some atomistic forcefields can already describe the interactions between the two phases with different degrees of accuracy\cite{Munetoh2007,Watanabe1999,Kulkarni2013}, a more general and robust framework is needed. Indeed, since Si and SiO$_2$ continue to be essential materials\cite{Kim2009,Wang2003,Neumayer2001,Son2015}, and since details at the nanoscale play an increasingly critical role in advanced electronic devices, a unified description of these multicomponent and heterogeneous systems is required focusing on interface phenomena and not only on bulk-like features. With the persistent increase in computational power, we are no longer limited to simple forcefields, and the relentless shrinking of microelectronics devices will inexorably require a fully-atomistic description of entire devices for their design, manufacturing, and reliability. From these ideas emerged the Charge-Optimized Many-Body (COMB) forcefield framework, aimed at deriving a general forcefield capable of describing covalent, metallic, ionic, and van der Waals interactions\cite{Liang2013}. Two distinct parameterization of previous generations of COMB already exist for Si-O interactions\cite{Yu2007,Shan2010}. The third generation of COMB supersedes previous versions, extending the reach of the forcefield to the interaction between organic molecules and metallic compounds\cite{Liang2012}. Parameters for various heterogeneous systems have been introduced. Here, we present a parameterization of COMB3 for Si/SiO$_2$ obtained exclusively using data from {\it ab initio} computations. The new parameterization is validated by both its fidelity to the {\it ab initio} training set and its ability to predict properties ranging from density to lattice thermal conductivity. 

The subsequent parts of this paper are organized as follows: section \ref{s-2} gives a short presentation of the functional form of COMB3. In section \ref{s-3}, we discuss in detail the full optimization procedure we have established. Section \ref{s-4} consists of a series of applications which serve as validation. We first evaluate how well our forcefield reproduces the structure and the energetics of crystalline SiO$_2$ polymorphs. We then extend the simulations to surface energies, vibrational density of states, lattice thermal conductivity, and finally the interfacial thermal resistance at the SiO$_2$/Si interface. Section \ref{s-5} provides a summary and conclusions. 

\section{Third generation COMB forcefield formalism}\label{s-2}

The third generation charge-optimized many-body forcefield aims at providing a robust and general formalism capable of modeling simultaneously the different types of chemical bonds that can be found in nature. A detailed description of the full functional form of COMB3 can be found in previously published papers\cite{Liang2013,Liang2012}. Here we clarify the components of the total potential energy function employed in the present study, which can be written as: 

\begin{equation}
\begin{split}
U^{tot}[\{q\},\{r\}] = &\quad U^{es}[\{q\},\{r\}] + U^{short}[\{q\},\{r\}] \\ 
&\quad + U^{vdW}[\{r\}] + U^{corr}[\{q\},\{r\}]
\end{split}
\label{eq-1}
\end{equation}

$U^{es}$ being the electrostatic energies, $U^{short}$ the short-range interactions, $U^{vdW}$ the van der Waals interactions, and $U^{corr}$ a set of correction terms. 

\subsection{Electrostatic terms}

The electrostatic energies include charge-charge interactions, charge-nuclear interactions, a self-energy term representing the energy to form a charge on each atom, and atomic polarizability: 

\begin{equation}
\begin{split}
U^{es}[\{q\},\{r\}] = &\quad U^{qq}[\{q\},\{r\}] + U^{qZ}[\{q\},\{r\}] \\ 
&\quad + U^{self}[\{q\}] + U^{pol}[\{r\}]
\end{split}
\label{eq-2}
\end{equation}

In this parameterization, the atomic polarizability, the correction and the van der Waals terms are set to zero, hence we will not detail their formalism here. The atomic charges are described with the use of Streitz-Mintmire\cite{Streitz1994} charge density distribution functions, to avoid the Coulombic catastrophe. The charge density of an atom is given as: 

\begin{equation}
\rho_i\left(\vec{r},q_i\right) = Z_i \delta \left( \left| \vec{r} - \vec{r_i} \right| \right) + \left( q_i - Z_i \right)f_i \left( \left| \vec{r} - \vec{r_i} \right| \right)
\label{eq-3}
\end{equation}

With $\vec{r}$ the spatial position, $\vec{r_i}$ the position of atom $i$, $Z_i$ an effective point core charge, here treated as a fitted parameter, $\delta$ the Dirac delta function, and $f(\vec{r})$ a function representing the radial decay of the electron density of the s-type orbital: 

\begin{equation}
f\left( \left| \vec{r} - \vec{r_i} \right| \right) = \zeta_i^3 \pi^{-1} \exp \left( -2 \zeta_i \left| \vec{r} - \vec{r_i} \right| \right)
\label{eq-4}
\end{equation}

$\zeta_i$, called the Slater orbital exponent, is a parameter controlling how stiff the decay is. The charge-charge and charge-nuclear interactions can be written as: 

\begin{equation}
\begin{split}
&U^{qq}[\{q\},\{r\}] = \sum_i \sum_{j>i} q_i J_{ij}^{qq} q_j \\
&U^{qZ}[\{q\},\{r\}] = \sum_i \sum_{j>i} \left( q_i J_{ij}^{qZ} Z_j + q_j J_{ji}^{qZ} Z_i \right)
\end{split}
\label{eq-5}
\end{equation}

Where $J_{ij}^{qq}$ is the Coulomb integral operator, and $J_{ij}^{qZ}$ the charge-nuclear coupling operator. The energy required to form a charge on an atom is approximated as a Taylor expansion series with respect to its charge, as stated by Mortier {\it et al.}\cite{Mortier1985} plus a correction term: 

\begin{equation}
\begin{split}
U^{self}[\{q\}] = \sum_i &\quad \left[ \vphantom{\sum_i} \chi_i q_i + J_i q_i^2 + K_i q_i^3 + L_i q_i^4 \right. \\
&\quad \left. \vphantom{\sum_i} + 100 \cdot \left( q_i - q_i^{lim} \right) q_i^4 \right] + V_i^{field}
\end{split}
\label{eq-6}
\end{equation}

Where $\chi_i$ and $J_i$ are respectively associated to an atom's electronegativity and atomic hardness. The correction term $V_i^{field}$ represents the change of electronegativity and atomic hardness of the atom due to its environment, using four adjustable parameters: 

\begin{equation}
\begin{split}
V_i^{field} = \frac{1}{4\pi\epsilon_0} 
\sum_{j > i} 
\left( \vphantom{\sum_i} \right. & \frac{P_{ij}^{\chi}q_j}{r_{ij}^3+\left(A_{ij}^{\chi}/r_{ij}\right)^3} \\
& \left. + \frac{P_{ij}^{J}q_j^2}{r_{ij}^5+\left(A_{ij}^{J}/r_{ij}\right)^5} \right)
\end{split}
\label{eq-7}
\end{equation}

\subsection{Short-range charge-dependent terms}

The bond energy is described with pairwise attractive and repulsive terms, both distance and charge dependent. The attractive part of the potential is coupled to a bond-order term. 

\begin{equation}
\begin{split}
U^{short}[\{q\},\{r\}] = 
\sum_i \sum_{j > i} & F_c(r_{ij}) \left[ \vphantom{\sum_i} V^R(r_{ij},q_{i,j}) \right. \\
& \left. \vphantom{\sum_i} - b_{ij} V^A(r_{ij},q_{i,j}) \right]
\end{split}
\label{eq-8}
\end{equation}

With $V^R$ and $V^A$ respectively the repulsive and attractive terms, $b_{ij}$ the bond-order term, and $F_c$ a Tersoff type cutoff function. We have: 

\begin{equation}
V^R = A_{ij} \exp \left[ - \lambda_{ij} r_{ij} + \lambda_i^* \right]
\label{eq-9}
\end{equation}

\begin{equation}
V^A = B_{ij}^* \exp \left( \alpha_i^* \right) \sum_{n=1}^3 \left[ B_{ij}^n \exp \left(- \alpha_{ij}^n r_{ij} \right) \right]
\label{eq-10}
\end{equation}

$\lambda_i^*$, $\alpha_i^*$ and $B_{ij}^*$ are charge-dependent functions, as initially derived by Yasukawa\cite{Yasukawa1996}. The attractive part uses three different exponentials: the flexibility it gives is required to describe carbon bonding. Only one exponential, and therefore two parameters instead of six are used for Si-O interactions. 

The bond-order term alters the short-range attraction between two atoms according to the local environment: 

\begin{equation}
b_{ij} = \frac{1}{2} \left( b_{ij}^{\sigma - \pi} + b_{ji}^{\sigma - \pi} \right) + b_{ij}^{\pi}
\label{eq-11}
\end{equation}

$b_{ij}^{\pi}$ captures the non-local conjugation effects in organic materials. It is decomposed into two contributions, one from the radical character, and the other one from dihedral angles. It is presently set to zero. $b_{ij}^{\sigma - \pi}$ is the bond-order contribution arising from covalent bonding: 

\begin{equation}
\begin{split}
b_{ij}^{\sigma - \pi} = 
& \left\{ 1 + \left[ \sum_{k \neq i,j}^{NN} F_c (r_{ik}) \zeta (r_{ij},r_{ik}) g_{ij}(\cos(\theta_{ijk})) \right]^{\eta_i} \right. \\
& \left. \vphantom{\sum_{kj}^{N}} + P_{ij}(\Omega_i)^{\eta_i} \right\}
\end{split}
\label{eq-12}
\end{equation}

\begin{equation}
\zeta (r_{ij},r_{ik}) = \exp \left[ \beta_{ij}^{m_i} \left( r_{ij} - r_{ik} \right)^{m_i} \right]
\label{eq-13}
\end{equation}

\begin{equation}
g_{ij}(x) = \sum_{n=0}^6 b_{n,ij} x^n
\label{eq-14}
\end{equation}

\begin{equation}
P_{ij}(\Omega_i) = c_0 \Omega_i + c_1 e^{c_2 \Omega_i} + c_3
\label{eq-15}
\end{equation}

The so-called symmetry function $\zeta (r_{ij},r_{ik})$ weakens longer bonds, like a screening function. Its strength is controlled by the parameter $\beta_{ij}$. The angular function $g_{ij}(\cos(\theta_{ijk}))$ is a sixth-order polynomial, which makes use of two-dimensional bond-specific parameters, $b_{1,ij}$ to $b_{6,ij}$. In most of the cases, the coordination function $P_{ij}(\Omega_i)$ has the analytical form presented in equation \ref{eq-15}. $\Omega_i$ is the coordination number of the central atom $i$, minus the atom $j$. For carbon-based bonds, where the dependence on the coordination number is complex, a tricubic spline replaces the analytical expression.

\section{Optimization procedure}\label{s-3}

This section explains the determination of the various parameters of this forcefield. First, we present the training set obtained from {\it ab initio} calculations. Then, we detail the fitting scheme. Finally, we discuss the resulting parameters, and probe the fidelity of the new forcefield with respect to the initial {\it ab initio} data. 

\subsection{{\it Ab initio} calculations}

All {\it ab initio} calculations were performed using density functional theory\cite{Hohenberg1964,Kohn1965} (DFT) as implemented in the Vienna {\it Ab initio} Simulation Package\cite{Kresse1993,Kresse1996} (VASP) and integrated in the MedeA$^\circledR$ computational environment\cite{MedeA}. Exchange-correlation effects were described by the  generalized gradient approximation (GGA) in the form of PBEsol proposed by Perdew et al\cite{Perdew2008}, which is a revised Perdew-Burke-Ernzerhof GGA\cite{Perdew1996} improving equilibrium properties of densely-packed solids and their surfaces. The Kohn-Sham equations are solved with the all-electron frozen-core projector augmented wave method\cite{Blöchl1994}, using plane-wave basis sets with a cutoff of 400 eV. 

In order to cover the largest field of possible configurations, we have included three different classes of systems in the training set: bulk structures, surfaces, and clusters of SiO$_2$. Depending on the system considered, we have extracted the fitting data from three different types of computations, namely {\it ab initio} molecular dynamics trajectories (AIMD), single point energy calculations (SPE), and structure optimizations (SO). The entire set of systems is presented in table \ref{table-1}. 

\begin{table}[h]
\caption{Bulk, surfaces, and clusters of SiO$_2$ considered in the training-set. E, F, S stand for Energy, Force vector, and Stress tensor. }
\centering
\begin{tabular}{c c c c}

 \hline\hline

 System & Symmetry & Calculation & Fitting data \\ [1ex]
 
 \hline
 
 \multicolumn{4}{ c }{SiO$_2$ bulks} \\ [1ex]
 alpha-quartz & {\it P3$_1$21} & AIMD & E, F, S \\
 alpha-cristobalite & {\it P4$_1$2$_1$2} & AIMD & E, F, S \\
 stishovite & {\it P4$_2$/mnm} & SPE & E, F, S \\
 keatite & {\it P4$_3$2$_1$} & SPE & E, F, S \\
 amorphous & {\it P1} & SPE & E, F, S \\ [0.5ex]
 
 \hline
 
 \multicolumn{4}{ c }{Alpha-quartz surfaces} \\ [1ex] 
 (0001) & {\it P1} & AIMD & E, F \\
 (100) & {\it P1} & SO & E, F \\
 (111) & {\it P1} & SO & E, F \\ [0.5ex]
 
 \hline
 
 \multicolumn{4}{ c }{SiO$_2$ clusters} \\ [1ex] 
 dimer & D$_{2h}$ & SO & E, F \\
 trimer & D$_{3h}$ & SO & E, F \\
 trimer & D$_{2d}$ & SO & E, F \\
 tetramer 1 & D$_{2h}$ & SO & E, F \\
 tetramer 2 & D$_{2h}$ & SO & E, F \\
 tetramer & C$_{2v}$ & SO & E, F \\
 tetramer & D$_{4h}$ & SO & E, F \\ [0.5ex]
 
 \hline\hline
 
\end{tabular} 
\label{table-1}
\end{table}

Crystalline bulk structures were obtained as follows. First, the unit cell was optimized by minimizing the energy and the forces with a conjugate gradient method. The convergence criterion on the forces was 0.02 eV/\AA, and the energy cutoff was set to 520 eV. Then, a supercell was built with the optimized unit cell, and used as an input for AIMD or SPE calculations, depending on the system considered. 

In order to include a description of disordered systems in our training-set, we have generated a sample of amorphous silica using classical molecular dynamics, together with the BKS forcefield\cite{vanBeest1990}, slightly adapted for molten SiO$_2$ using a stiff repulsive core. A time step of 1 fs has been used. Starting from a crystalline structure of beta-cristobalite, the system was heated from 300 K up to 5000 K for 100 ps, in the isothermal-isobaric ({\it NPT}) ensemble. The melt was then kept at 5000 K for 200 ps, in the canonical ({\it NVT}) ensemble. After that, the system was quenched from 5000 K to 300 K, during 500 ps. In a final step, the resulting structure was relaxed in the {\it NPT} ensemble, until the density was converged under standard thermodynamic conditions. An {\it ab initio} SPE calculation was then performed on the structure with the same settings as for crystalline systems. 

We have included surfaces and small clusters of SiO$_2$ in the training-set, in order to have information on under-coordinated Si and O atoms, which is not available with crystalline systems. We have selected seven oligomers with low oligomerization degree n, going from two to four. The structures are stable (SiO$_2$)$_n$ systems described by Harkless et al.\cite{Harkless1996}, using the TTAM forcefield\cite{Tsuneyuki1988}. Structural optimization was used to sample a large number of configurations, starting with slightly distorted structures. Three different alpha-quartz surfaces were also included in the training-set. AIMD and SO were performed on the surface samples, by freezing several layers of the interior of the system, thus allowing only the atoms close to the surface to move. 

\subsection{The fitting procedure}

The fitting process was decomposed into two stages. Starting with an initial set of parameters, a) the parameter space was extensively probed using a genetic algorithm b) the resulting parameters were optimized with a non-linear least-squares solver. During the first step of optimization, we introduced separate objective functions, through Pareto optimization. Considering a list of observables $\{ \theta \}$, which can be a list of energies, forces or stress components, one can write: 

\begin{equation}
\theta^{MD} = a \theta^{AI} + b
\label{eq-16}
\end{equation}

Where MD stands for molecular dynamics, and AI for {\it ab initio}. During the multi-objective optimization step, one can choose to use the slope $a$, the intercept $b$ and the regression coefficient $R^2$ of the linear function in addition to the usual Root Mean Square Deviation (RMSD) as separate objectives. The ideal values for $a$, $b$, and $R^2$ are respectively 1, 0, and 1. When switching from multi-objective to mono-objective optimization, the final Pareto front was considered. One set of parameters, $\left\{ \eta_i^P \right\}$, was selected as follows: 

\begin{equation}
\left\{ \eta_i^P \right\} = 
\max_{PF} \left( \sum_{j}^{n_{OF}} S_j \right)
\label{eq-17}
\end{equation}

The $\max$ function runs on all the points of the Pareto front (PF). $S_j$ is the normalized score of a given set of parameters for a given objective function (OF). $n_{OF}$ is the total number of objective functions. 

We employ a standard non-linear least squares procedure to solve the following function: 

\begin{equation}
\Gamma \left( \left\{ \eta_i \right\} \right) = w_E \Delta E + w_F \Delta F + w_S \Delta S
\label{eq-18}
\end{equation}

with respect to the parameters $\left\{ \eta_i \right\}$, where

\begin{equation}
\begin{split}
& \Delta E = \frac{\sqrt{
\sum_{c=1}^{N_c}
\left| E_{c,i}^{MD} - E_{c,i}^{AI} \right|^2}
}
{\sqrt{
\sum_{c=1}^{N_c}
\left( E_{c,i}^{AI} \right)^2
}} \\[1em]
& \Delta F = \frac{\sqrt{
\sum_{c=1}^{N_c}\sum_{i=1}^{N}\sum_{\alpha}^{\alpha = x, y, z}
\left| F_{c,i}^{MD,\alpha} - F_{c,i}^{AI,\alpha} \right|^2
}}
{\sqrt{
\sum_{c=1}^{N_c}\sum_{i=1}^{N}\sum_{\alpha}^{\alpha = x, y, z}
\left( F_{c,i}^{AI,\alpha} \right)^2
}} \\[1em]
& \Delta S = \frac{\sqrt{
\sum_{c=1}^{N_c}\sum_{i=1}^{N}\sum_{\beta,\gamma}^{\beta,\gamma = x, y, z}
\left| S_{c,i}^{MD,\beta\gamma} - S_{c,i}^{AI,\beta\gamma} \right|^2
}}
{\sqrt{
\sum_{c=1}^{N_c}\sum_{i=1}^{N}\sum_{\beta,\gamma}^{\beta,\gamma = x, y, z}
\left( S_{c,i}^{AI,\beta\gamma} \right)^2
}}
\end{split}
\label{eq-19}
\end{equation}

$w_{E,F,S}$  are adjustable weights on the energies, forces, and stress. $N_c$ is the total number of configurations, and $N$ the total number of atoms in a given configuration. $\alpha$,  $\beta$, and $\gamma$ represent the different components of force vectors and stress tensors. 

When using total energies from VASP calculations in the fit of COMB3 parameters, one needs to ensure that the same reference for isolated neutral atoms is used. In the case of a forcefield such as COMB3, this energy is zero by definition. In VASP calculations, the total energy is referred to isolated, spherical, non spin-polarized atoms (without multiplet splitting) computed with the corresponding PAW potential. Hence, for open-shell atoms a shift is required, which is the VASP energy of spin-polarized atoms without restrictions to spherical symmetry. The values of this shift are 1.5756 eV for oxygen and 0.8224 eV for silicon. 

\subsection{The new forcefield}

We have used the original parameters for O-O interactions\cite{Liang2012} to ensure compatibility with previous parameterization of COMB3, though it should be noted that it was found crucial to reduce the O-O cutoff distance, from $2.4-2.8$ \AA, to $1.8 - 2.2$ \AA. This modification is physically meaningful, since O-O bonds are usually much shorter than the latter values. In vitreous silica, the O-O distance is 2.65 \AA \cite{Mozzi1969}. It was found that using a cutoff value higher than the experimental O-O distance strongly affected computed properties, through the bond-order term. Using high cutoff values for O-O interactions is however possible, but requires a re-parameterization of the O-O terms. Here, we chose to ensure the highest compatibility with other COMB3 parameterizations. This modification will only affect systems including O-O bonds, like ozone, dioxygen, or peroxides. However, it should be noted that COMB3 has never been parameterized for such compounds. 

Trying to optimize all parameters of COMB3 simultaneously is impractical, because (i) the parameters space is rather large, and (ii) different terms from different parts of the functional form are correlated. Hence, a one-step automatic optimization will strip the physical meaning of the different terms of the forcefield. Therefore, we proceeded in five distinct steps: 

\begin{enumerate}
\item{Setting initial parameter values. Short-range and bond-order parameters were derived from the SiO$_2$ Tersoff parameterization by Munetoh {\it et al.}\cite{Munetoh2007}, together with COMB2. The other parameters were set to reasonable values derived from other parameterizations of COMB3\cite{Liang2012}. }
\item{Deriving the Taylor expansion self-energy parameters directly from quantum chemistry calculations using TURBOMOLE\cite{Furche2014}, at the CCSD (T) level of theory, and with QZVPP\cite{Weigend2003} basis-sets. }
\item{Optimizing all other parameters using the full training-set, with respect to energies and forces. }
\item{Re-optimizing the four short-range parameters using only the bulk structures, fitting the parameters to energies and stress. }
\item{Re-optimizing the pure Si short-range parameters using a super-cell of diamond Si at equilibrium, fitting the parameters to energies and stress. }
\end{enumerate}

Once the optimization cycle is completed, the performances of the resulting forcefield with respect to the original data are evaluated. The four final criteria are computed ($a$, $b$, $R^2$, and the RMSD), as well as a plot of the reference observables versus the computed ones, called a AI/MD graph. 

\begin{figure*}
\begin{center}
  \includegraphics[height=8cm]{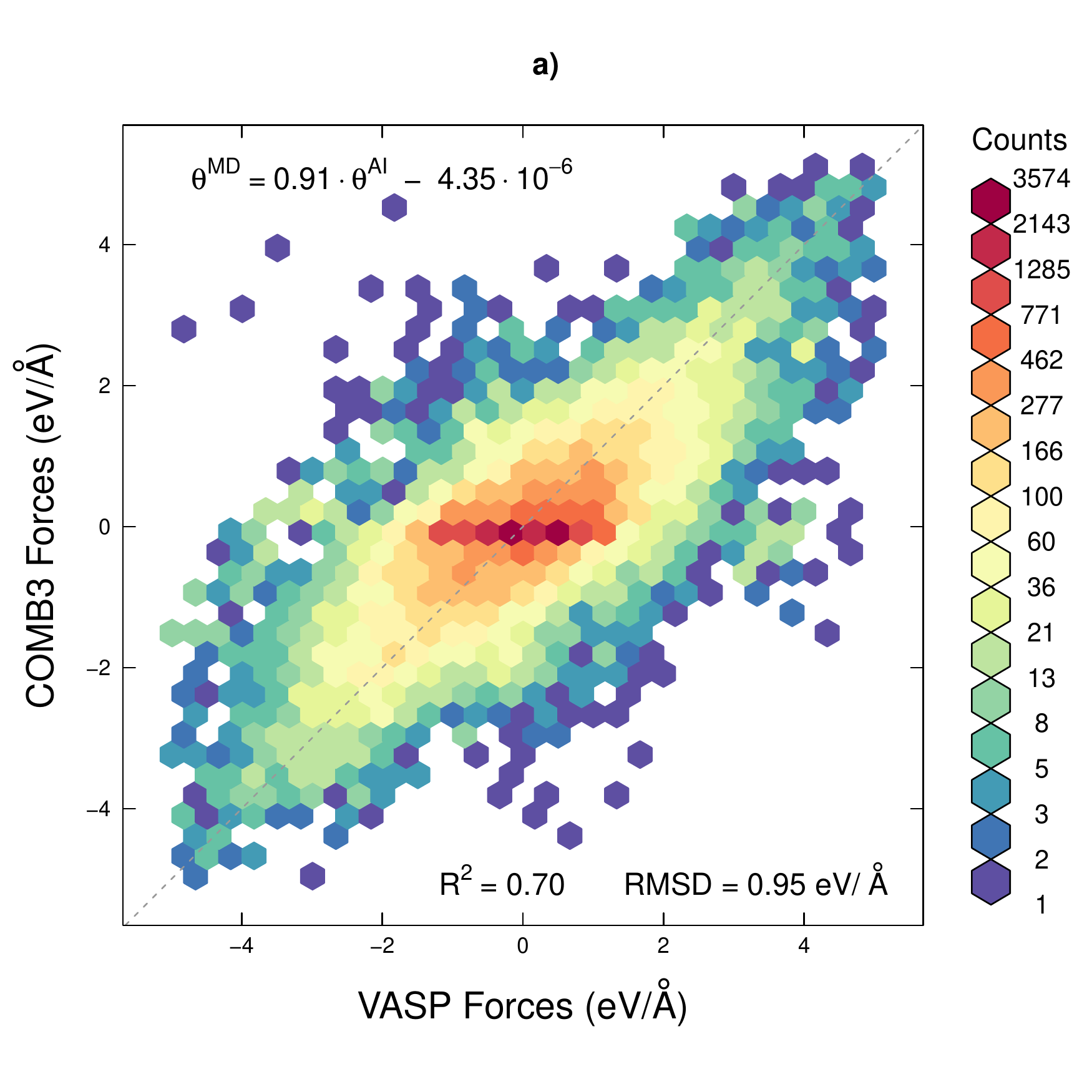}
  \includegraphics[height=8cm]{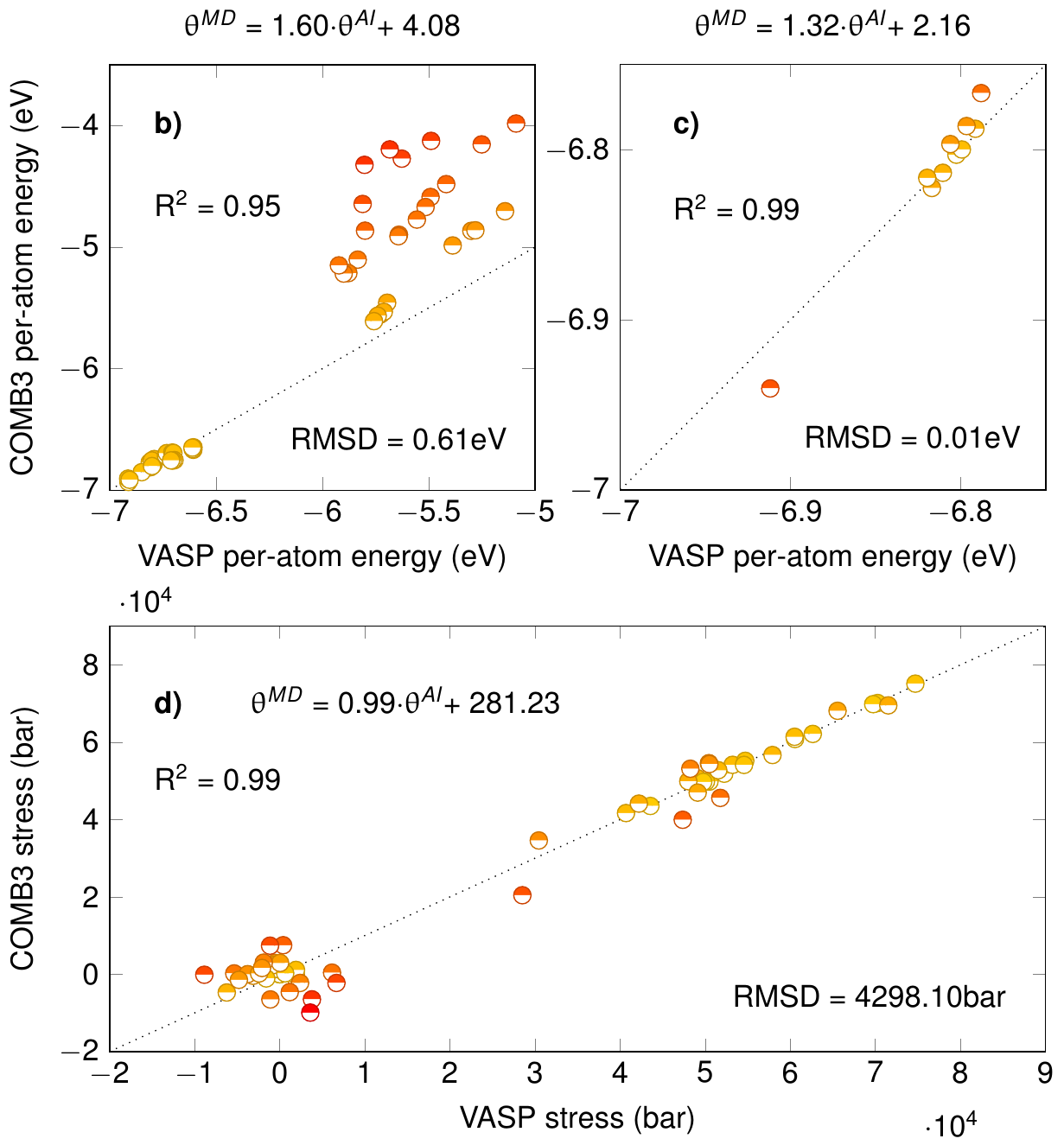}
  \caption{AI/MD plots of the two main optimization stages. a) Forces comparison for stage 3, colors represent the number of data points in each hexagon. b) and c) Per-atom energy for stages 3 and 4, respectively. d) Stress comparison for stage 4. On each plot, the gray dashed line represents the $\theta^{MD}=\theta^{AI}$ function. Plots b), c), and d) are colored according to the deviation. }
  \label{fig-1}
\end{center}
\end{figure*}

Figure \ref{fig-1} shows four AI/MD graphs from optimization stages 3 and 4. Two of the graphs present per-atom energies and the two others, forces and stress components. The four criteria are also shown on the graphs. As a general behavior, the per-atom energies of oligomers are overestimated. On the other hand, bulk systems and surfaces are very well described. All the parameter values are presented in tables \ref{table-2} and \ref{table-3}. 

\begin{table}
\caption{COMB3 atom-type parameters for SiO$_2$. }
\centering
\begin{tabular}{l l l}

 \hline\hline

 Parameters & Si & O \\ [1ex]
 \hline
 
 $\chi$ & 1.1998236556 & 6.59963 \\
 $J$ & 5.7795364627 & 5.955097 \\
 $K$ & 0 & 0.7604334 \\
 $L$ & 0 & 0.009388015 \\
 $\zeta$ & 1.993643 & 1.371794 \\
 $Z$ & 2.125065 & -1.53917 \\
 $D_L$ & 1.187223 & 0.007664409 \\
 $D_U$ & -1.30496 & -1.213951 \\
 $Q_L$ & -4 & -2 \\
 $Q_U$ & 4 & 6 \\
 $q_{min}$ & -4 & -2 \\
 $q_{max}$ & 4 & 6 \\
 $\lambda^*$ & 0.08992715 & 5.295119 \\
 $\alpha^*$ & 0.5105194 & 3.258854 \\
 $m$ & 1 & 1 \\
 $n$ & 0.78734 & 1 \\
 \hline\hline
 
\end{tabular} 
\label{table-2}
\end{table}

\begin{table}
\caption{COMB3 bond-type parameters for SiO$_2$. }
\centering
\begin{tabular}{l l l l l}

 \hline\hline

 Parameters & Si-Si & Si-O & O-Si & O-O \\ [1ex]
 \hline
 $\lambda$ & 2.606093 & 3.1032401 & 3.1032401 & 5.295119 \\
 $A$ & 2390.192 & 1378.5063 & 1378.5063 & 4956.339 \\
 $\alpha$ & 1.769593 & 2.046672 & 2.046672 & 3.258854 \\
 $B$ & 513.247 & 450.0 & 450.0 & 688.1635 \\
 $\beta$ & 0 & 0 & 0 & 3.258854 \\
 $b_0$ & 0.057289 & 0.057289 & 1.169784 & 0.8565567 \\
 $b_1$ & 0.191259 & 0.191259 & 2.653276 & 1.826597 \\
 $b_2$ & 0.15898 & 0.15898 & 1.313542 & -0.2046884 \\
 $b_3$ & -0.001448 & -0.001448 & -0.275894 & -5.652039 \\
 $b_4$ & -0.000597 & -0.000597 & -0.048044 & 1.257097 \\
 $b_5$ & 0 & 0 & 0 & 16.00164 \\
 $b_6$ & 0 & 0 & 0 & 14.17783 \\
 $c_0$ & 0 & -0.04954035 & 0.8964422 & 0 \\
 $c_1$ & 0 & -0.01393876 & 0.02585726 & 0 \\
 $c_2$ & 0 & -1 & -1 & 0 \\
 $c_3$ & 0 & 0.1869605 & 0.0002308561 & 0 \\
 $P^{\chi}$ & -1.226449 & 0.001252494 & 0.8603318 & 1.966411 \\
 $P^{J}$ & 2 & 2 & 0.06534864 & 2.521788 \\
 $A^{\xi}$ & 0.5 & 0.5 & 0.5 & 0.25 \\
 $A^{J}$ & 0.5 & 0.5 & 0.5 & 0.25 \\
 $R$ & 2.5 & 2.06 & 2.06 & 1.8 \\
 $S$ & 2.8 & 2.37 & 2.37 & 2.2 \\
 \hline\hline
 
\end{tabular} 
\label{table-3}
\end{table}

\section{Results}\label{s-4}

Given that the new forcefield exhibits satisfying statistical characteristics concerning the fidelity of reproducing the values of the training set, we now compute various properties to assess its quality, and to define its domain of applicability. 

\subsection{Structure and per-atom energy}

We investigate first the density, structural aspects, and energetics of nine SiO$_2$ polymorphs. The density at finite temperature was evaluated for every system by performing molecular dynamics in the {\it NPT} ensemble, at thermodynamic conditions where the material is stable or metastable. Most simulations were therefore performed at standard conditions ($T_0 = 298.15$ K, $P_0 = 1$ atm), except for beta-phases, which are unstable at this temperature. We changed the equilibrium temperature for the three phases: 1000 K for $\beta$-quartz, 1300 K for $\beta$-tridymite, and 1700 K for $\beta$-cristobalite. Charge equilibration was performed and the density ($\rho^T$) was sampled during 10 ps, once the equilibrium was reached. We also computed the density ($\rho_0$) and per-atom energy ($E_0$) at zero temperature, in order to compare with our {\it ab initio} data. We computed energy-volume curves, considering only isotropic deformations without minimizing the energy, because minimization sometimes caused phases to change. For example, optimizing $\beta$-quartz systems leads to a breaking of the hexagonal symmetry in favor of the trigonal crystal system of $\alpha$-quartz. 

In order to get a good comparison of the present forcefield with other methods, molecular dynamics calculations using other forcefields have been performed. We chose four forcefields, which all have a different formalism: a Born-Mayer-Buckingham potential, namely BKS\cite{vanBeest1990}, COMB2\cite{Shan2010}, the Tersoff parametrization from Munetoh et al. \cite{Munetoh2007}, and ReaxFF\cite{Kulkarni2013}. All the computations were done with LAMMPS, using reax/c\cite{Aktulga2012} for ReaxFF, which is its c++ implementation. 

The results are listed in table \ref{table-4} and table \ref{table-5}. Zero temperature density should be directly compared to the density obtained from {\it ab initio} calculations. One can see that the agreement is excellent, a result which gives confidence in the fidelity of the forcefield to reproduce DFT values. The finite temperature density is to be compared to experimental data, and the agreement between the two is also excellent for most of the polymorphs. Keatite, and $\beta$-quartz yield information on range of applicability of the forcefield. The computed density obtained for keatite at standard conditions is $\sim 7\%$ higher than experiment, but its structural features are adequately represented. For $\beta$-quartz, the density is also high ($\sim 5\%$) and its configuration relaxes to a structure similar to $\alpha$-quartz, sharing its trigonal symmetry (symmetry constraints are not employed in these calculations). Hence we find that the forcefield possesses limitations in predicting phase stability.

There are slight differences in the cohesive energy of the different polymorphs of SiO$_2$ of the order of 1 meV/atom. These subtleties are in general not captured correctly with classical molecular dynamics, except in cases where the forcefield has been specially fitted to reproduce such small differences, which then can lead to larger deviations in other situations. One should first notice that the RMSD on the reduced training set, presented in figure \ref{fig-1}, is 10 meV/atom - an order of magnitude higher than the variations in cohesive energy. $\alpha$-quartz is the most stable phase of silicon dioxide, and hence is considered as the ground state. Here, we compare results obtained with DFT and classical molecular dynamics, using different forcefields. The PBEsol functional itself does not reproduce correctly the ordering in SiO$_2$ polymorphs cohesive energy: $\alpha$-cristobalite is obtained as the most stable phase, followed by $\alpha$-tridymite. Hybrid DFT, using for example the HSE06\cite{Heyd2003} functional in conjunction with PBEsol, gives the correct ordering. However, its computational cost is currently prohibitive, particularly where molecular dynamics trajectories are required as in the present application. With that in mind, one can question if such a high level of theory is necessary, given that the RMSD of the fidelity of the forcefield to the training-set will most certainly be larger than the required precision. Our parametrization of COMB3 does not give alpha-quartz as the ground state, coesite being computed to be slightly more stable. However, the general trend is correct: alpha phases are more stable than high temperature beta phases, and stishovite, having a rutile symmetry, is significantly less stable. The differences in cohesive energy between different polymorphs are somewhat out of scale: this is a general trend of all forcefields we have considered. Only two forcefields give the correct ordering: Tersoff and COMB2. The ordering of polymorphs was taken as an explicit criterion during the fitting process of the latter one, using correction functions to fine-tune it. Nonetheless, the equilibrium density of SiO$_2$ polymorphs predicted by these two forcefields severely deviate from the experimental data. It should also be noted that the stability of a phase is determined by the Gibbs free energy, while the above considerations are based solely on the total energy. Hence, some deviation from the experimentally observed ranking may be due to this simplification.

\begin{table*}
\caption{Density and per-atom energies of nine SiO$_2$ polymorphs as computed with COMB3}
\centering
\begin{tabular}{l c c c c c r}

 \hline\hline

 Polymorph & Symmetry & $a_0$, $b_0$, $c_0$ (\AA) & q$_{Si}$ ($e$) & $\rho^T$ (g/cm$^3$) & $\rho_{0}$ (g/cm$^3$) & $E_0$ (eV/atom) \\ [1ex]
 \hline
 
 $\alpha$-quartz & {\it P3$_1$21} & 4.965, 5.464 & 1.017 & 2.659 & 2.566 & -6.9626 \\
 $\beta$-quartz & {\it P6$_2$22} & 5.048, 5.537 & 1.013 & 2.657 & 2.449 & +0.0251 \\
 $\alpha$-cristobalite & {\it P4$_1$2$_1$2} & 4.997, 6.956 & 1.006 & 2.398 & 2.298 & +0.0226 \\
 $\beta$-cristobalite & {\it Fd-3m} & 7.438 & 1.005 & 2.200 & 1.940 & +0.1413 \\
 $\alpha$-tridymite (MX-1) & {\it Cc} & 5.112, 8.782, 8.389 & 1.002 & 2.319 & 2.120 & +0.0691\\
 $\beta$-tridymite & {\it P6$_3$/mmc} & 5.260, 8.590 & 1.005 & 2.226 & 1.939 & +0.1408 \\
 coesite & {\it C2/c} & 7.144, 12.326, 7.118 & 1.033 & 2.964 & 2.938 & -0.0132 \\
 keatite & {\it P4$_3$2$_1$} &  7.509, 8.666 & 1.009 & 2.672 & 2.450 & +0.0454 \\
 stishovite & {\it P4$_2$/mnm} & 4.286, 2.741 & 1.040 & 3.987 & 3.963 & +0.4729 \\[0.5ex]
 
 \hline\hline
 
\end{tabular} 
\label{table-4}
\end{table*}

\begin{table*}
\caption{Density and per-atom energies of nine SiO$_2$ polymorphs coming from experimental results, {\it ab initio} calculations and with other forcefields}
\centering
\begin{tabular}{l c | c c c c c | r r r r r }

 \hline\hline

 Polymorph & \multicolumn{1}{ c }{$\rho_0$} & \multicolumn{5}{ c }{$\rho^T$ (g/cm$^3$)} & \multicolumn{5}{ c }{$E_a$ (eV/atom)} \\ [1ex]
 \hline
  & DFT & Expt. & BKS & COMB2 & ReaxFF & Tersoff & DFT & BKS & COMB2 & ReaxFF & Tersoff \\
 $\alpha$-quartz & 2.577 & 2.649\cite{Wright1981} & 2.560 & 2.832 & 2.586 & 2.446 & -8.2222 & -19.4027 & -6.9204 & -7.6716 & -6.6978 \\
 $\beta$-quartz & 2.402 & 2.534\cite{Wright1981} & 2.472 & 2.733 & 2.599 & 2.357 & +0.0060 & +0.0388 & +0.0854 & -0.0195 & +0.0140 \\
 $\alpha$-cristobalite & 2.264 & 2.332\cite{Downs1994} & 2.406 & 2.721 & 2.202 & 2.137 & -0.0020 & +0.0978 & +0.0838 & -0.0058 & +0.0008 \\
 $\beta$-cristobalite & 1.942 & 2.175\cite{Barth1932} & 2.193 & 2.421 & 2.240 & 2.012 & +0.0107 & +0.1035 & +0.2662 & -0.0172 & +0.1410 \\
 $\alpha$-tridymite (MX-1) & 2.138 & 2.310\cite{Xiao1995} & 2.445 & 2.533 & 2.202 & 2.069 & -0.0009 & +0.0364 & +0.1615 & -0.0085 & +0.0100 \\
 $\beta$-tridymite & 1.941 & 2.211\cite{Villars1991} & 2.312 & 2.476 & 2.229 & 2.032 & +0.0079 & +0.0885 & +0.2742 & -0.0149 & +0.0024 \\
 coesite & 2.861 & 2.918\cite{Levien1981} & 2.845 & 2.956 & 2.903 & 2.733 & +0.0172 & -0.0073 & +0.0042 & +0.0135 & +0.0145 \\
 keatite & 2.398 & 2.499\cite{Shropshire1959} & 2.506 & 2.769 & 2.589 & 2.322 & +0.0060 & +0.0044 & +0.1108 & +0.0014 & +0.0127 \\
 stishovite & 4.229 & 4.283\cite{Sugiyama1987} & 4.326 & 3.330 & 2.767 & 3.907 & +0.0582 & -0.0121 & +1.0883 & +1.5508 & +0.4884 \\[0.5ex]
 
 \hline\hline
 
\end{tabular} 
\label{table-5}
\end{table*}

As a conclusion, we can stress that the present parameterization of COMB3 gives a satisfactory description of the different polymorphs of SiO$_2$. Densities are very close to experimental and {\it ab initio} values, and the energetic ordering of the phases is reasonably well reproduced. 

\subsection{\texorpdfstring{$\alpha$-quartz surface energy}{alpha-quartz surface energy}}

The variable charge scheme included in COMB3 enables accurate predictions where the distribution of charges is anisotropic, as in surfaces and interfaces. Here, we have computed the surface energy of three different states of $\alpha$-quartz's (0001) surface. The cleaved surface is known to be highly reactive, due to dangling bonds. We have investigated two reconstruction patterns which are thought to be the most stable forms\cite{Chen2008}, namely a ($1 \times 1$) dense pattern and a ($2 \times 1$) reconstruction. In order to rapidly obtain reconstructed systems, we have followed the procedure outlined by Chen et al.\cite{Chen2008}. Starting with the oxygen-terminated freshly cleaved (0001) surface of $\alpha$-quartz, we have performed simulated annealing up to 1400 K, at a rate of 50 K/ps, using the BKS forcefield. By extracting structures at different temperatures during the annealing, we were able to obtain both ($1 \times 1$) and ($2 \times 1$) surface reconstruction patterns. The different systems were then annealed at low temperature and had their potential energy minimized, using our new parameterization and BKS. The surface energy $E_S$ of the minimized structures is then defined as: 

\begin{equation}
E_S = \frac{E - N E_a}{2 A}
\label{Senergy}
\end{equation}

\begin{figure}
\begin{center}
  \includegraphics[width=8cm]{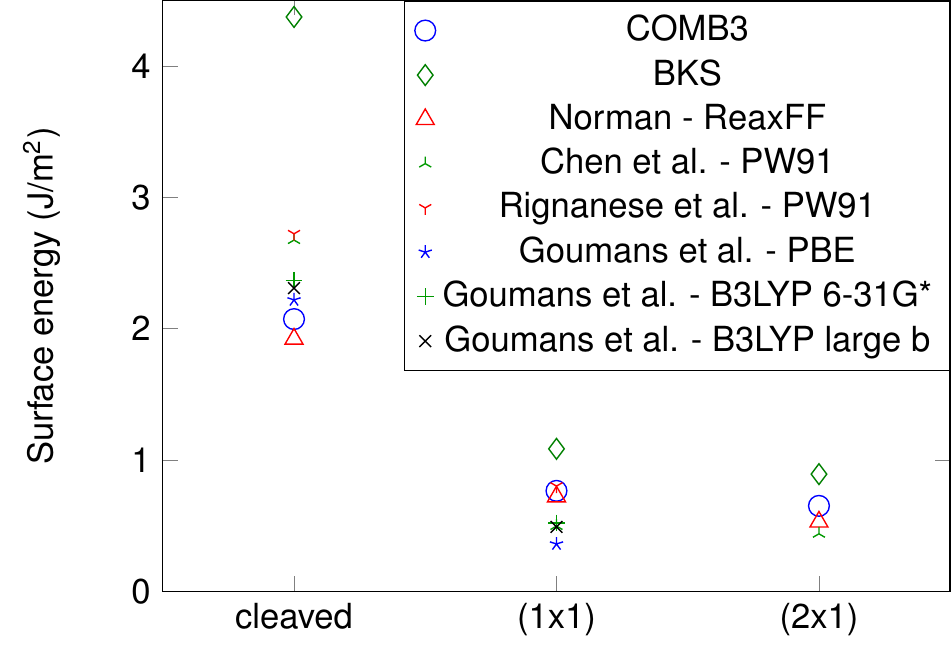}
  \caption{Surface energy of alpha-quartz (0001) surface, considering three models: the freshly cleaved surface, and two reconstruction patterns. Different methods are compared: DFT-GGA\cite{Goumans2007,Chen2008,Rignanese2004}, hybrid DFT\cite{Goumans2007}, ReaxFF\cite{Norman2013}, BKS, and COMB3. }
  \label{fig-2}
\end{center}
\end{figure}

With $N$ the number of atoms in the system, $E_a$ the cohesive energy, and $A$ the surface area. We also compare, on figure \ref{fig-2}, our findings to different {\it ab initio} studies\cite{Goumans2007,Chen2008,Rignanese2004} and to results obtained using ReaxFF\cite{Norman2013}. 

The surface energy predicted by COMB3 is in good agreement with {\it ab initio} data. The cleaved, highly reactive surface, has a surface energy of 2.07 J/m$^2$, which compares well to the value of 2.22 J/m$^2$ obtained using the PBE functional. The overall first-principles results on reconstructed surfaces are well represented by the forcefield. ReaxFF behaves comparably well similar to the present parameterization of COMB3. BKS overestimates the surface energy of the cleaved surface, leading to a highly reactive system. Reconstruction immediately occurs, even during potential energy minimization only. However, the global trend of the surfaces stability is well reproduced. 

\subsection{Dynamical properties}

In order to further validate our parameterization, we have computed dynamical properties of both ordered and disordered phases. First, we have evaluated the vibrational density of states of both $\alpha$-quartz and amorphous SiO$_2$, at room temperature. We have also computed the lattice thermal conductivity of $\alpha$-quartz and of amorphous silica, which is the essential contribution to the total thermal conductivity in insulators or semiconductors. 

Evaluation of the vibrational density of states was achieved by the following steps. We first sampled the atomic positions during a 10 ps long run in the micro-canonical ({\it NVE}) ensemble, at room temperature. It was then possible to assemble the position autocorrelation function, and to obtain the vibrational density of states by direct application of the Wiener-Khinchine theorem: 

\begin{equation}
G(\nu) = \frac{1}{N k_B T} \sum_j \int_{- \infty}^{\infty} e^{2 \pi i \nu t} \frac{\langle \vec{r_j}(t) \cdot \vec{r_j}(0)\rangle}{\langle \vec{r_j}(0) \cdot \vec{r_j}(0)\rangle} dt
\label{vdos}
\end{equation}

\begin{figure}
\begin{center}
  \includegraphics[width=8cm]{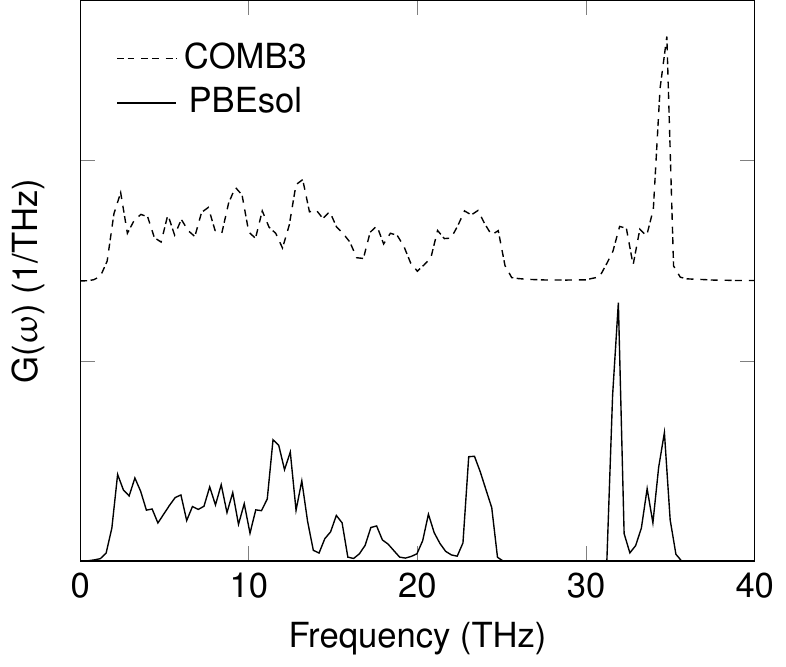}
  \caption{Comparison of the vibrational density of states of alpha-quartz computed using DFT and the optimized forcefield. }
  \label{fig-3}
\end{center}
\end{figure}

\begin{figure}
\begin{center}
  \includegraphics[width=8cm]{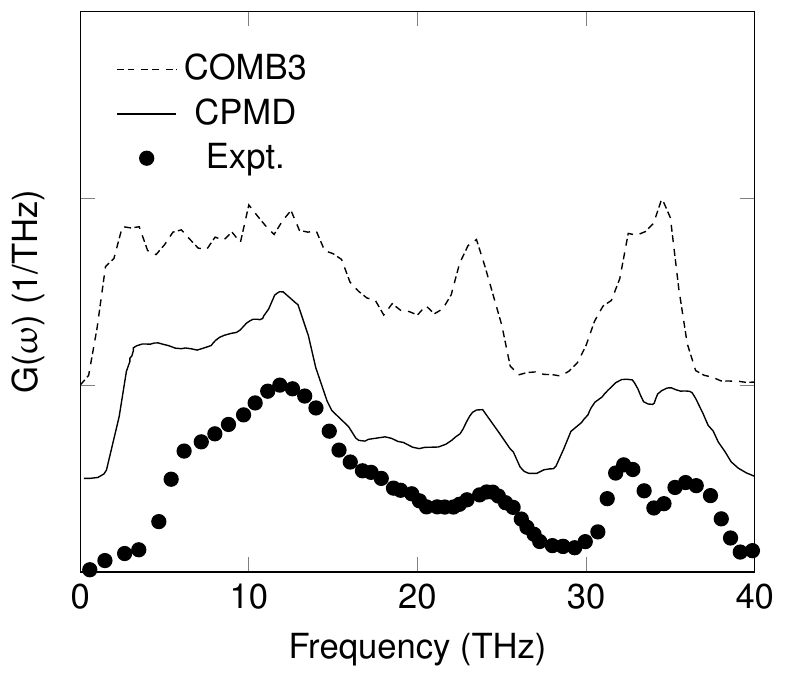}
  \caption{Comparison of the experimental\cite{Carpenter1985} vibrational density of states of amorphous silica, to the computed VDOS using our forcefield and CPMD\cite{Rahmani2003}. }
  \label{fig-4}
\end{center}
\end{figure}

Where $N$ is the number of atoms, $k_B$ the Boltzmann constant, and $\vec{r_j} (t)$ the position of atom $j$ at time $t$. We have averaged our results over three different samples for amorphous silica. As a reference, we have computed the vibrational density of states with the PBEsol functional using the MedeA-Phonon code\cite{Parlinski1997}. The results are compared in figures \ref{fig-3} and \ref{fig-4}. The computed vibrational density of states $G(\omega)$ of amorphous silica is compared to experimental results using inelastic neutron scattering\cite{Carpenter1985}, and to a theoretical calculation\cite{Rahmani2003} using Car-Parrinello molecular dynamics (CPMD). 

The vibrational density of states computed with COMB3 is in very good agreement with that obtained from {\it ab initio} calculations. The lower-frequency vibrations, the so-called rigid-unit modes (RUM) which are responsible of inter-tetrahedral motions, are well-reproduced. The observed peaks at 11.4 THz and 23.1 THz are also correctly described. At higher frequencies, a double peak structure originates from intra-tetrahedral stretching. At 34.6 THz, the four oxygen atoms in a SiO$_4$ tetrahedron move relative to the central silicon atom. At 32 THz, two oxygen atoms vibrate in anti-phase with one another. The position of these peaks is also very precisely reproduced, although their intensities are somewhat swapped. 

The main features of the vibrational density of states of amorphous silica are also very well reproduced. Here, we have normalized each set of results using the value of its maximum peak, to allow direct comparison. The experimental density of states presents two peaks at approximately 12 and 24 THz, respectively associated with rocking and bending modes, which are accurately reproduced. The double peak structure at $32 - 36$ THz, corresponding to stretching modes, is slightly too populated, but the peak positions are correct. The low-frequencies peak at ca. 3 THz, here experimentally invisible because of the insufficient resolution\cite{Wischnewski1998}, is very well described. 

We have chosen to probe the behavior of the thermal conductivity of alpha-quartz in the [0001] direction as a function of the temperature, in order to make a direct comparison to experimental measurements\cite{Kanamori1968}. The thermal conductivity of the amorphous phase at 300 K has also been computed. Reverse non-equilibrium molecular dynamics\cite{Mullerplathe1997} (RNEMD) has been used where a heat flux is applied to the system, and the resulting temperature gradient is measured. The thermal conductivity is then obtained using Fourier's law: 

\begin{equation}
J_{\mu} = - \sum_{\nu} \kappa_{\mu \nu} \partial T / \partial x_{\nu}
\label{fourier}
\end{equation}

\begin{figure}
\begin{center}
  \includegraphics[width=8cm]{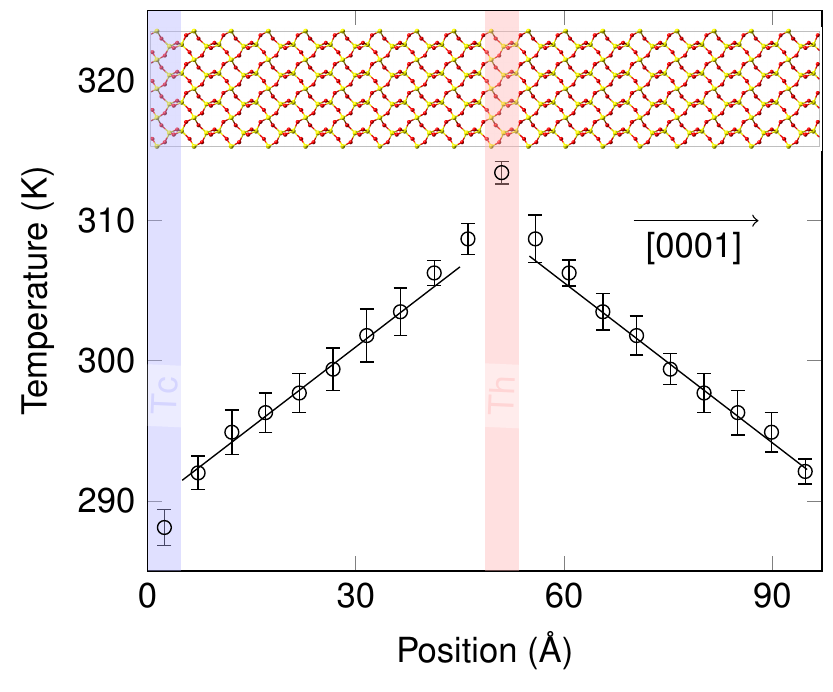}
  \caption{Usual setup for an RNEMD calculation: the system is divided in bins, and a temperature gradient is established between two thermostat bins. }
  \label{fig-5}
\end{center}
\end{figure}

Where $J_{\mu}$ is an element of the heat flux density vector, $\kappa_{\mu \nu}$ is an element of the lattice thermal conductivity tensor, and $\partial T / \partial x_{\nu}$ is the temperature gradient in the $\nu$ direction. The system was first equilibrated at the desired temperature in the {\it NPT} ensemble. Once the equilibrium volume was converged at an average pressure of 1 atm, the system was equilibrated in the {\it NVT} ensemble for 500 ps. Subsequently, RNEMD was performed. To this end, the system was divided into $n$ slabs perpendicular to the heat flux direction, which here is along the optical axis. During 1 or 2 ns in the {\it NVE} ensemble, depending on the convergence rate of the temperature gradient, the highest and lowest atomic kinetic energies, respectively from the first and the $1+n/2$ layers, were swapped at a specific rate. This induces a temperature gradient in the system which is subsequently measured. The whole {\it modus operandi} is presented in figure \ref{fig-5}. 

Finite size effects strongly affect the thermal conductivity computed by means of non-equilibrium molecular-dynamics, when the sample considered is smaller than the mean free path of heat carriers in an infinite system. The regime where the thermal conductivity is limited by the system size is known as the Casimir limit. As a consequence, we extrapolate the thermal conductivity at infinite system size through linear regression by evaluating the thermal conductivity as a function of the inverse of the system size in the transport direction. Considering three or four different system sizes is usually enough to obtain a high-enough linear regression coefficient. 

\begin{figure}
\begin{center}
  \includegraphics[width=8cm]{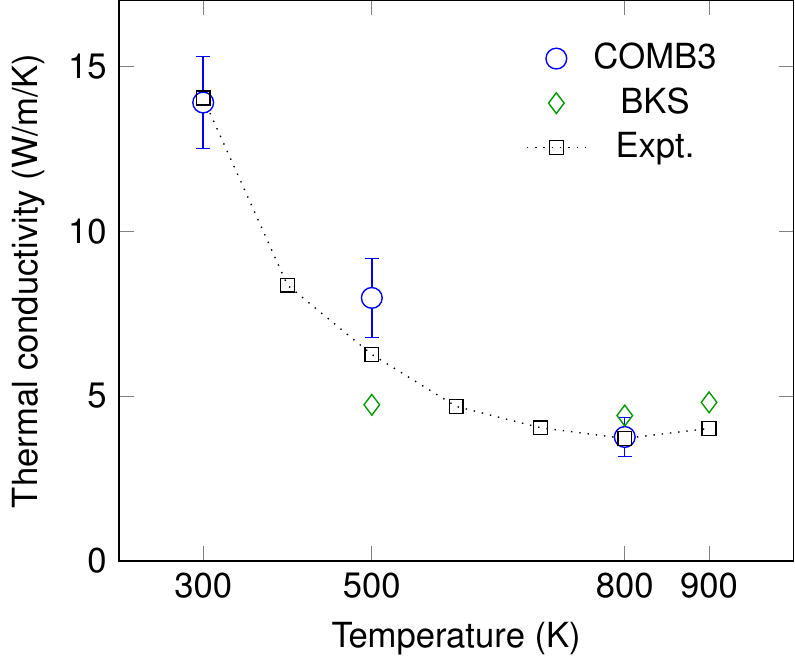}
  \caption{Thermal conductivity in the [0001] direction of $\alpha$-quartz as a function of temperature, as computed with COMB3, with BKS\cite{Yoon2004}, and compared to experimental results\cite{Kanamori1968}. Note the very good agreement between COMB3 and the experimental results. }
  \label{fig-6}
\end{center}
\end{figure}

The thermal conductivity of $\alpha$-quartz along the optical axis as a function of the temperature is presented on figure \ref{fig-6}. It is compared to experimental results, where the thermal conductivity is estimated from the measured thermal diffusivity, and from non-equilibrium molecular dynamics simulations using BKS. The agreement with experimental data is very good. The thermal conductivity tends to decrease in the $300 - 800$ K temperature range, due to increased inelastic scattering or Umklapp processes. This trend is well reproduced here. At 300 K and at 800 K, our calculations are even in perfect agreement with the experiment. Unfortunately, there are no data points for the 300 K - 400 K range in the study of Yoon {\it et al.} \cite{Yoon2004} with BKS. The computations are in fairly good agreement with the experiment. However, the overall trend doesn't seem correct, which could be due to uncertainties. 

For amorphous silica, we have obtained a thermal conductivity at room temperature of $\kappa = 1.455$ W/m/K, with an uncertainty of $\pm 0.044$ W/m/K. The experimental values range from 1.3 to 1.5 W/m/K\cite{Regner2013,Cahill1988,Lee1997,Yamane2002}. There has been a number of other studies reporting computed values with other forcefields. In a recent study, Larkin\cite{Larkin2014} reports a value of around 2.1 W/m/K using BKS and equilibrium molecular dynamics. Shenogin reported a similar value of 2.0 W/m/K\cite{Shenogin2009} using the same forcefield and NEMD. Yeo\cite{Yeo2013}, with means of RNEMD, reports a value of 2.13 W/m/K using BKS, and 1.19 W/m/K using Tersoff. The BKS forcefield is known to fail in reproducing the vibrational properties of amorphous silica\cite{Benoit2002}, especially at low frequencies. Here, it is shown that the COMB3 forcefield provides a thermal conductivity of amorphous SiO$_2$ at room temperature in good agreement with experimental observation. 

\subsection{Thermal transport across c-Si/a-SiO$_2$ interface}

Encouraged by the accuracy of the new forcefield parametrization in predicting structural, energetic, vibrational, and thermal conductivity of pure phases, we now present results of the thermal conductivity across a semiconductor/oxide interface. As stated earlier, COMB3 is well suited for describing heterogeneous structures, because of the variable charge approach and its general functional form. We have computed the interface thermal resistance, also known as the Kapitza resistance, at the interface between crystalline silicon and amorphous silica. This interface exists in almost every Si-based microelectronic device like transistors, where the thermal management is of growing concern for feature sizes in the sub 10 nm scale. It is therefore essential to be able to correctly model this interface at the atomistic scale, in order to understand the underlying mechanisms of heat transport. 

As structural model we have chosen a periodic system consisting of a 16a$_0$-thick layer of amorphous silica embedded in crystalline Si with a$_0$ being the lattice constant of Si in the diamond structure. The entire system is 30 nm long, and contains two interfaces. As previously outlined by Chen et al. \cite{Chen2012}, the microscopic topology of the c-Si/a-SiO$_2$ interface highly influences the Kapitza resistance. We have therefore computed an average using both interfaces of the system. In order to match amorphous silica and crystalline silicon, we have prepared several amorphous SiO$_2$ bulk systems using a melt and quench process. The sample with the smallest proportional lattice mismatch in the $(xy)$ plane with c-Si was then selected, in order to minimize strain in the resulting interface system. The amorphous part was then matched with a crystalline silicon sample. An initial gap of 1.5 \AA was added at the interface to prevent overlapping atoms. The whole system was then annealed in the {\it NPT} ensemble at a temperature of 2000 K during a 20 ps run at ambient pressure including charge optimization. The resulting interfaces are flat and atomically sharp, which is probably not what one can expect of an interface between an oxide and a semiconductor. Experimentally, a suboxide layer with varying stoichiometry up to 1.6 nm thick\cite{Devine1996} is observed. In order to obtain more realistic systems, we have annealed one of our samples at 1500 K during 500 ps, charges being optimized every 100 fs. During the annealing, oxygen atoms diffuse in the crystalline Si, inducing disorder. We thus obtained diffuse interfaces with suboxide layers about 1 nm thick. Both systems have been used to compute the Kapitza resistance R$_k$ in the heat transport direction $\mu$, which can be defined as: 

\begin{equation}
R_{k} = \frac{\Delta T_{\mu}}{J_{\mu}}
\label{kapitza}
\end{equation}

\begin{figure}
\begin{center}
  \includegraphics[width=8cm]{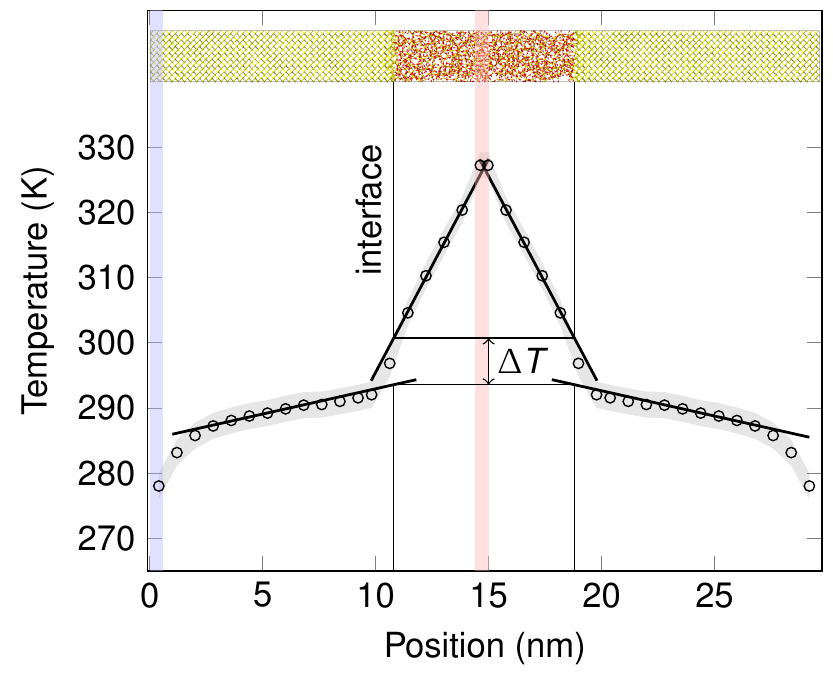}
  \caption{Model for determining the Kapitza resistance.}
  \label{fig-7}
\end{center}
\end{figure}

\begin{figure}
\begin{center}
  \includegraphics[width=8cm]{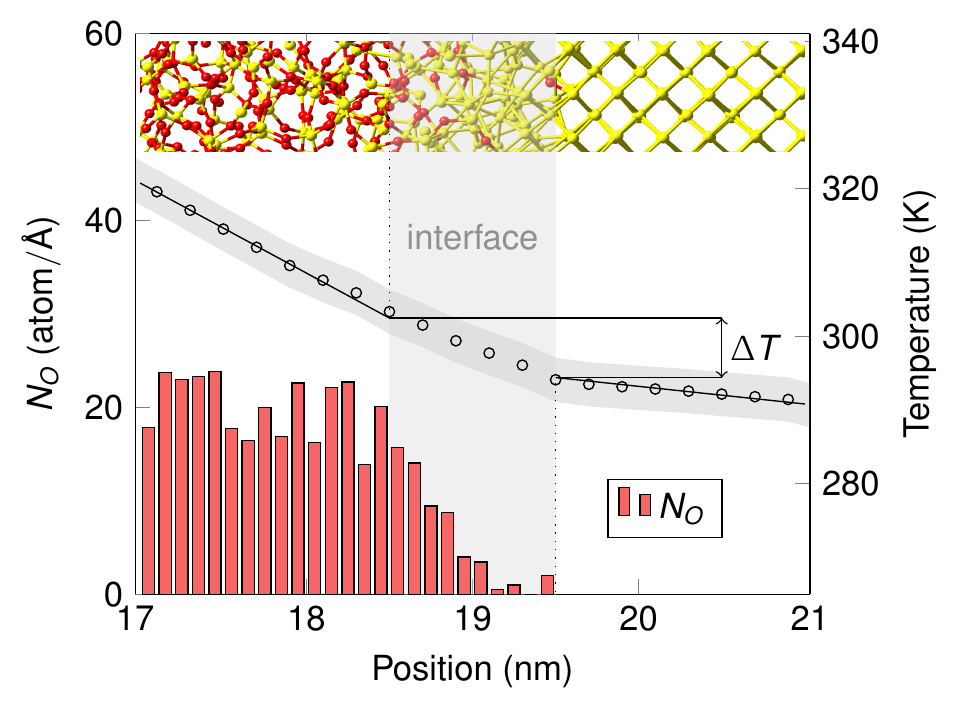}
  \caption{Detail of the structure and temperature gradient at the interface, in the annealed system. The temperature drop is taken as the difference in temperature from both sides of the suboxide layer. The thickness of the suboxide layer is estimated by plotting the average oxygen concentration as a function of the position (red bars). }
  \label{fig-8}
\end{center}
\end{figure}

With $J_{\mu}$ the heat flux density in the transport direction, and $\Delta T_{\mu}$ the temperature gap at the interface. We can therefore use RNEMD to compute the interfacial thermal resistance, and the value of the temperature gap can be extrapolated by performing linear regressions of the temperature in regions where the gradient is linear, as shown in figure \ref{fig-7}. For the annealed system, the temperature difference is evaluated from both parts of the interface, as one can see in figure \ref{fig-8}. The thickness of the interface is estimated using the average oxygen concentration. 

\begin{figure}
\begin{center}
  \includegraphics[width=8cm]{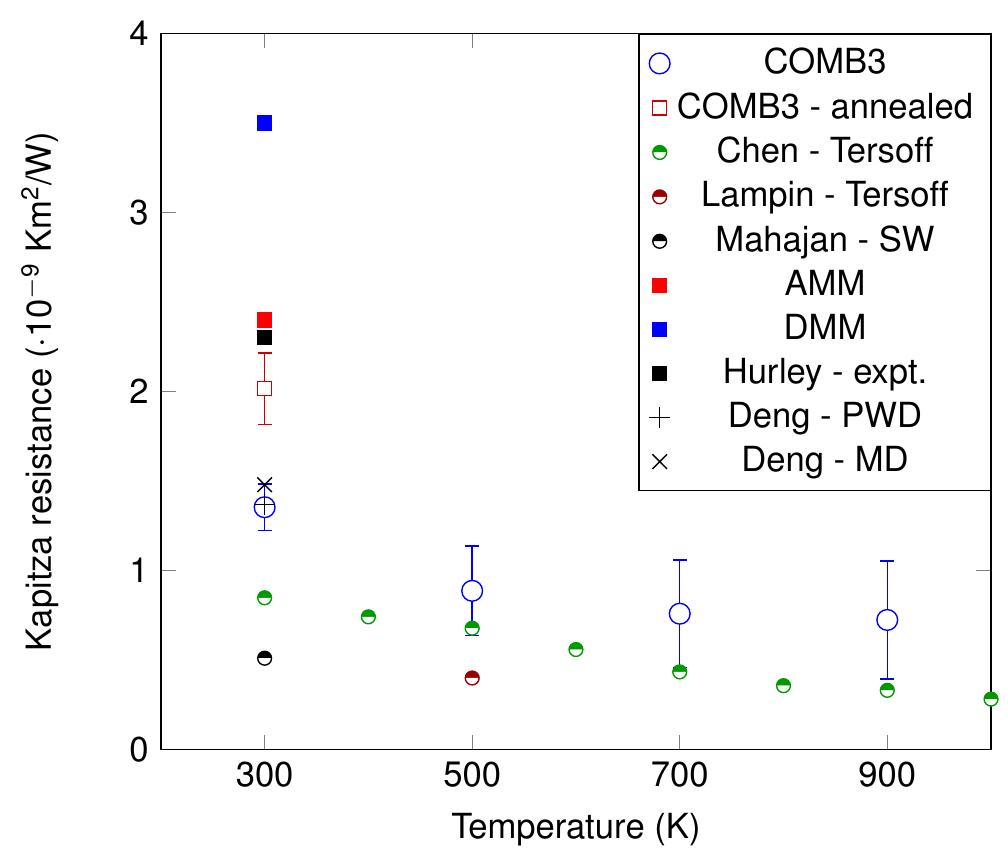}
  \caption{Kapitza resistance as a function of temperature computed using COMB3, and compared to experimental results, AMM and DMM theoretical predictions, and classical molecular dynamics studies. }
  \label{fig-9}
\end{center}
\end{figure}

We have computed the Kapitza resistance at several temperatures. For each point, the system was equilibrated in the {\it NPT} ensemble at the target temperature. It was then thermalized in the {\it NVT} ensemble for 500 ps. The RNEMD run itself was 1 ns up to 2 ns long, depending on the convergence of the temperature gradient. Our results are presented in figure \ref{fig-9}, along with previous studies.  

There have been several studies of the Kapitza resistance between a-SiO$_2$ and c-Si. Hurley {\it et al.} \cite{Hurley2011} measured the interfacial thermal resistance across a bicrystal interface of silicon, using time resolved thermal wave microscopy. They have observed the existence of a 4.5 nm thick layer of silica at the interface between the two crystals, and estimated a Kapitza resistance value of $2.3 \cdot 10^{-9}$ Km$^2$/W, using a continuum thermal transport model. Mahajan\cite{Mahajan2011}, in 2011, was the first to use molecular dynamics to give an estimate of the Kapitza resistance, with an extended Stillinger-Weber forcefield\cite{Watanabe1999}. Two studies published in 2012, from Lampin {\it et al.} \cite{Lampin2012} and Chen {\it et al.} \cite{Chen2012}, investigated the Kapitza resistance between silica and silicon using the Tersoff forcefield. Chen showed that when the coupling between Si and SiO$_2$ is strong enough, the Kapitza resistance does not depend on the thickness of the amorphous layer. Lampin devised a new method, called "approach-to-equilibrium" MD (AEMD), to probe thermal transport between two media. In their work, the Kapitza resistance was derived from an expression of the total thermal conductance as a sum of different contributions. Finally, in 2014, Deng {\it et al.}\cite{Deng2014} performed an extended study of the thermal transport at c-Si/a-SiO$_2$ interfaces at room temperature, using both NEMD and a phonon wave packet dynamics method (PWD). Deng {\it et al.} extracted the Kapitza resistance in the same way as Lampin, {\it i.e.} through linear regression of the total thermal conductance as a function of the amorphous layer thickness $l_A$, the interfacial thermal resistance being extrapolated at $l_A \rightarrow 0$. We have chosen a different route, as explained above, which avoids the need for multiple calculations for different values of $l_A$. 

At room temperature, our results agree very well with the computations of Deng {\it et al.} The experimental data is considered as an upper limit of the Kapitza resistance, since several phenomena affecting the thermal transport such as sample purity are not being taken into account in our atomistic computations. The behavior with temperature is well captured, and in agreement with Chen {\it et al.}'s findings: inelastic scattering increases with increasing temperature, hence lowering the value of the Kapitza resistance. However, the values of the present simulations are shifted towards the experimental values, a feature that we associate to the different forcefields. The Tersoff forcefield used by Chen is short-ranged, and may fail at capturing some of the thermal transport phenomena whereas the present COMB3 parametrization captures a large range of effects. The previous theoretical molecular dynamics studies have always considered atomically sharp interfaces. Here, by taking into account the existence of a suboxide layer, which one can consider as a more realistic model, we obtain results much closer to the experimental values. These results cannot be obtained using simpler forcefields as Tersoff or Stillinger-Weber, because they would not be capable of describing a complex phenomenon like oxygen diffusion in crystalline Si, which require an explicit description of long-range interactions, and charge transfer. This is one of the reasons why more complex forcefields such as COMB3 are necessary. 

\section{Conclusions}\label{s-5}

In this work, we have derived a new set of parameters for a charge-optimized many-body (COMB3) forcefield for systems containing Si/SiO$_2$ based solely on information from {\it ab initio} calculations. Given the large number of mutually dependent parameters in the COMB3 description, a step-wise optimization has been employed, starting with the most important short-range parameters and then successively including additional terms. Reactive forcefields such as COMB3 contain an aggressive cutoff procedure to discriminate between bonded and non-bonded neighbors. This aspect needs to be carefully handled as it can lead to numerical instabilities if the integration steps in molecular dynamics simulations are too large. Although highly automated procedures for the optimization of forcefield parameters have been employed in the present work, the somewhat large number of parameters in COMB3 makes the fitting procedure very delicate.

The new forcefield parameters have been applied and evaluated by computing (i) the structure and relative energies of nine different SiO$_2$ polymorphs, (ii) surface energies, (iii) vibrational densities of states of crystalline and amorphous silica, (iv) thermal conductivity of $\alpha$-quartz and amorphous silica, and (v) the Kapitza resistance of a c-Si/a-SiO$_2$ interface. Overall, the performance of the new parametrization is better than that of other forcefields reported so far in the literature. In particular, this new COMB3 forcefield is capable of predicting the correct surface stability of $\alpha$-quartz, and brilliantly reproduces the vibrational features of ordered and disordered systems. The Kapitza resistance of the c-Si/a-SiO$_2$ interfaces computed in this work is close to earlier simulations by Deng {\it et al.}, who reported only value for $T$ = 300 K. Values for higher temperatures obtained in the present work demonstrate a gentle decrease of the Kapitza resistance reaching  at 900 K about half the value of the resistance at 300 K. Contrary to simple forcefields without charge transfer and bond-order terms, the exhaustive formalism of COMB3 allows an accurate description of heterosystems, which can be of crucial importance in cases such as semiconductor/oxide interfaces with interdiffusion. Here, it is demonstrated that taking into account the suboxide layer between a semiconductor and an oxide yields interfacial thermal resistance values which are in much better agreement with the experiment. 

In conclusion it has to be stated that the mapping of a quantum mechanical many-body system onto a quasi-classical forcefield remains a delicate balance between accuracy and generality, especially when significant changes in local stoichiometry and topology are involved. However, once such a forcefield is established yielding realistic structural models and meaningful relative energies, the calculation of the thermal conductivity is actually less demanding on a forcefield as it needs to capture only the motion of each atom around its equilibrium position. The present work demonstrates that it is possible to parameterize a bond order potential with charge equilibration such that properties related to structure and energy as well as to thermal transport can be computed with high accuracy. This has been demonstrated for different polymorphs of silicon dioxide, amorphous SiO$_2$, oxide surfaces, and Si/SiO$_2$ heterostructures. The newly developed set of parameters is compatible with those reported by other researchers for systems containing C, O, and H, and thus expands the range of possible applications for the COMB3 forcefield. 


\begin{acknowledgments}
We gratefully thank Cray Inc., which has graciously provided most of the required computational time on a Cray XC-40 for the large-scale simulations of the present work. We also thank Beno{\^i}t Leblanc, who developed the optimization tools. The authors are also grateful to the ANRT (Association Nationale de la Recherche et des Technologies) and the MESR (Minist{\`e}re de l'Enseignement Sup{\'e}rieur et de la Recherche) for supporting this work through a CIFRE grant N$^\circ$2013/1403. 
\end{acknowledgments}

%
%


\bibliography{aps_bib}

%
%
%
%
%
%

\end{document}